\documentclass[italian,english]{article}
\usepackage[latin9]{inputenc}
\usepackage{color}
\usepackage{amsmath}
\usepackage{amssymb}

\newcommand{\lyxaddress}[1]{
\par {\raggedright #1
\vspace{1.4em}
\noindent\par}
}

\usepackage{babel}

\begin{document}

\title{\textbf{Will gravitational waves confirm Einstein's General Relativity?}}

\author{\textbf{Christian Corda }}

\maketitle

\lyxaddress{\begin{center}
Associazione Scientifica Galileo Galilei, Via Pier Cironi 16 - 59100
PRATO, Italy 
\par\end{center}}

\begin{center}
\textit{E-mail address:} \textcolor{blue}{cordac.galilei@gmail.com} 
\par\end{center}
\begin{abstract}
Even if Einstein's General Relativity achieved a great success and
overcame lots of experimental tests, it also showed some shortcomings
and flaws which today advise theorists to ask if it is the definitive
theory of gravity. In this proceeding paper it is shown that, if advanced
projects on the detection of Gravitational Waves (GWs) will improve
their sensitivity, allowing to perform a GWs astronomy, accurate angular
and frequency dependent response functions of interferometers for
GWs arising from various Theories of Gravity, i.e. General Relativity
and Extended Theories of Gravity, will be the ultimate test for General
Relativity. This proceeding paper is also a short review of the Essay
which won Honorable Mention at the 2009 Gravity Research Foundation
Awards.
\end{abstract}
Recently, the data analysis of interferometric GWs detectors has been
started (for the current status of GWs interferometers see \cite{key-1})
and the scientific community hopes in a first direct detection of
GWs in next years. 

Detectors for GWs will be important for a better knowledge of the
Universe and either to confirm or rule out the physical consistency
of General Relativity or of any other theory of gravitation \cite{key-2,key-3,key-4,key-5,key-6,key-7}.
In fact, in the context of Extended Theories of Gravity, some differences
between General Relativity and the others theories can be pointed
out starting by the linearized theory of gravity \cite{key-2,key-3,key-4,key-5,key-6,key-7,key-8,key-9,key-10,key-11,key-12}.
In this tapestry, detectors for GWs are, in principle, sensitive to
a hypothetical \textit{scalar} component of gravitational radiation
appearing in extended theories of gravity too, i.e. theories like
scalar-tensor gravity \cite{key-4,key-8,key-9,key-10,key-12}, bi-metric
theory \cite{key-5}, high order theories \cite{key-2,key-3,key-6,key-7},
Brans-Dicke theory \cite{key-13} and string theory \cite{key-14}.

Reasons of extending General Relativity arise from the fact that,
even if Einstein's Theory \cite{key-15} achieved a great success
(see for example the opinion of Landau who says that General Relativity
is, together with Quantum Field Theory, the best scientific theory
of all \cite{key-16}) and overcame lots of experimental tests \cite{key-15},
it also showed some shortcomings and flaws which today advise theorists
to ask if it is the definitive theory of gravity \cite{key-17,key-18}.
Differently from other field theories like the electromagnetic theory,
General Relativity has not been quantized. This fact rules out the
possibility of treating gravitation like other quantum theories, and
precludes the unification of gravity with other interactions. 

On the other hand, one can define \textit{Extended Theories of Gravity}
those semiclassical theories where the Lagrangian is modified, in
respect to the standard Einstein-Hilbert gravitational Lagrangian,
adding high-order terms in the curvature invariants (terms like $R^{2}$,
$R^{\alpha\beta}R_{\alpha\beta}$, $R^{\alpha\beta\gamma\delta}R_{\alpha\beta\gamma\delta}$,
$R\Box R$, $R\Box^{k}R$) or terms with scalar fields non minimally
coupled to geometry (terms like $\phi^{2}R$) \cite{key-17,key-18}.
In general, it is well known hat terms like those have to be considered
in all the approaches to perform the unification between gravity and
other interactions. More, from a cosmological point of view, such
modifies of General Relativity generate inflationary frameworks which
are very important as they solve lots of problems of the Standard
Universe Model \cite{key-19}. We emphasize that we are not telling
that General Relativity is wrong. It is well known that, even in the
context of Extended Theories, General Relativity remains the most
important part of the structure \cite{key-4,key-7,key-17,key-18}.
We are only trying to understand if weak modifies on such a structure
could be needed to solve some theoretical and observing problems \cite{key-17,key-18}.
We also recall that even Einstein told that General Relativity could
not be definitive \cite{key-28}. In fact, during his famous research
on the Unified Field Theory, he tried to realize a theory that he
called {}``Generalized Theory of Gravitation'', and he said that
 mathematical difficulties precluded him to obtain the final equations
\cite{key-28}.

In the general context of cosmological evidences, there are also other
considerations which suggest an extension of General Relativity. As
a matter of fact, the accelerated expansion of the Universe, which
is today observed, shows that cosmological dynamic is dominated by
the so called Dark Energy, which gives a large negative pressure.
This is the standard picture, in which such new ingredient is considered
as a source of the \textit{right side} of the field equations. It
should be some form of un-clustered non-zero vacuum energy which,
together with the clustered Dark Matter, drives the global dynamics.
This is the so called {}``concordance model'' ($\Lambda$CDM) which
gives, in agreement with the CMBR, LSS and SNeIa data, a good tapestry
of the today observed Universe, but presents several shortcomings
as the well known {}``coincidence'' and {}``cosmological constant''
problems \cite{key-20}. An alternative approach is changing the \textit{left
side} of the field equations, seeing if observed cosmic dynamics can
be achieved extending General Relativity \cite{key-17,key-18,key-21,key-22}.
In this different context, one does not require to find out candidates
for Dark Energy and Dark Matter, that, at the present time, have not
been found, but only the {}``observed'' ingredients, which are curvature
and baryon matter, have to be taken into account. Considering this
point of view, we can think that gravity is different at various scales
\cite{key-21} and a room for alternative theories is present. In
principle, the most popular Dark Energy and Dark Matter models can
be achieved considering $f(R)$ theories of gravity, where $R$ is
the Ricci curvature scalar, and/or Scalar-Tensor Gravity \cite{key-17,key-18,key-22}.

In this proceeding paper we show that, if advanced projects on the
detection of GWs will improve their sensitivity, allowing to perform
a GWs astronomy \cite{key-1}, accurate angular and frequency dependent
response functions of interferometers for GWs arising from various
Theories of Gravity, i.e. General Relativity and Extended Theories
of Gravity, will be the ultimate test for General Relativity \cite{key-2,key-3,key-4,key-5,key-6,key-7,key-10,key-11,key-12}.
The papers which found this essay have been the world's most cited
in the official Astroparticle Publication Review of ASPERA during
the 2007 with 13 citations \cite{key-23}. We recall that ASPERA is
the network of national government agencies responsible for coordinating
and funding national research efforts in Astroparticle Physics, see
\cite{key-30}. This proceeding paper is also a short review of the
Essay which won Honorable Mention at the 2009 Gravity Research Foundation
Awards \cite{key-29}.

Working with $G=1$, $c=1$ and $\hbar=1$ (natural units), the line
element for a GW arising from standard General Relativity and propagating
in the $z$ direction is \cite{key-1,key-15,key-24,key-25} 

\begin{equation}
ds^{2}=dt^{2}-dz^{2}-(1+h_{+})dx^{2}-(1-h_{+})dy^{2}-2h_{\times}dxdy,\label{eq: metrica TT totale}\end{equation}

where $h_{+}(t+z)$ and $h_{\times}(t+z)$ are the weak perturbations
due to the $+$ and the $\times$ polarizations which are expressed
in terms of synchronous coordinates in the Transverse Traceless (TT)
gauge \cite{key-15}. In \cite{key-24,key-25} it has been shown that
the total frequency and angular dependent response function (i.e.
the detector pattern) to the $+$ polarization of an interferometer
with arms in the $u$ and $v$ directions in respect to the propagating
GW is:\begin{align}
\tilde{H}^{+}(\omega) & \equiv\Upsilon_{u}^{+}(\omega)-\Upsilon_{v}^{+}(\omega)\nonumber \\
 & =\frac{(\cos^{2}\theta\cos^{2}\phi-\sin^{2}\phi)}{2L}\tilde{H}_{u}(\omega,\theta,\phi)-\frac{(\cos^{2}\theta\sin^{2}\phi-\cos^{2}\phi)}{2L}\tilde{H}_{v}(\omega,\theta,\phi),\label{eq: risposta totale Virgo +}\end{align}

that, in the low frequencies limit ($\omega\rightarrow0$) gives the
well known low frequency response function of \cite{key-26,key-27}
for the $+$ polarization: 

\begin{equation}
\tilde{H}^{+}(\omega)=\frac{1}{2}(1+\cos^{2}\theta)\cos2\phi+O\left(\omega\right)\,.\label{eq: risposta totale approssimata}\end{equation}

The derivation of eq. (\ref{eq: risposta totale Virgo +}) has been
shown in \cite{key-24,key-25,key-29} using the {}``bouncing photons
analysis'' that was created in \cite{key-31}. Actually, this kind
of analysis has strongly generalized to angular dependences, scalar
waves and massive GWs in \cite{key-2,key-4,key-12,key-24,key-25,key-29}.

In the same way, the response function for the $\times$ polarization
has been obtained as \cite{key-24,key-25,key-29}

\begin{equation}
\tilde{H}^{\times}(\omega)=\frac{-\cos\theta\cos\phi\sin\phi}{L}[\tilde{H}_{u}(\omega,\theta,\phi)+\tilde{H}_{v}(\omega,\theta,\phi)],\label{eq: risposta totale Virgo per}\end{equation}
that, in the low frequencies limit ($\omega\rightarrow0$), gives
the low frequency response function of \cite{key-26,key-27} for the
$\times$ polarization: \begin{equation}
\tilde{H}^{\times}(\omega)=-\cos\theta\sin2\phi+O\left(\omega\right)\,.\label{eq: risposta totale approssimata 2}\end{equation}

The case of massless Scalar-Tensor Gravity has been discussed in \cite{key-4,key-12}
with a {}``bouncing photons analysis'' similar to the previous one.
In this case, the line-element in the TT gauge can be extended with
one more polarization, labelled with $\Phi(t+z)$, i.e.

\begin{equation}
ds^{2}=dt^{2}-dz^{2}-(1+h_{+}+\Phi)dx^{2}-(1-h_{+}+\Phi)dy^{2}-2h_{\times}dxdy.\label{eq: metrica TT super totale}\end{equation}

The total frequency and angular dependent response function of an
interferometer to this {}``scalar'' polarization is \cite{key-4,key-12}\begin{align}
\tilde{H}^{\Phi}(\omega) & =\frac{\sin\theta}{2i\omega L}\{\cos\phi[1+\exp(2i\omega L)-2\exp i\omega L(1+\sin\theta\cos\phi)]+\nonumber \\
 & -\sin\phi[1+\exp(2i\omega L)-2\exp i\omega L(1+\sin\theta\sin\phi)]\}\,,\label{eq: risposta totale Virgo scalar}\end{align}

that, in the low frequencies limit ($\omega\rightarrow0$), gives
the low frequency response function of \cite{key-9,key-14} for the
$\Phi$ polarization: \textbf{\begin{equation}
\tilde{H}^{\Phi}(\omega)=-\sin^{2}\theta\cos2\phi+O(\omega).\label{eq: risposta totale approssimata scalar}\end{equation}
}

In \cite{key-2,key-3,key-4,key-7} it has also been shown that, in
the framework of GWs, the cases of massive Scalar-Tensor Gravity and
$f(R)$ theories are totally equivalent. It is well known that there
is a more general conformal equivalence between Scalar-Tensor Gravity
and $f(R)$ theories, even if there is a large debate on the possibility
that such a conformal equivalence should be also an effective \emph{physical}
equivalence \cite{key-17,key-18,key-21}. In such cases, because of
the presence of a small mass, a longitudinal component is present
in the third polarization. This implies that it is impossible to extend
the TT gauge to the third mode \cite{key-2,key-3,key-4,key-6,key-7}.
But gauge transformations permit to put the line-element due to such
a third scalar mode in a conformally flat form \cite{key-2,key-3,key-4,key-6,key-7}: 

\begin{equation}
ds^{2}=[1+\Phi(t-v_{G}z)](-dt^{2}+dz^{2}+dx^{2}+dy^{2}).\label{eq: metrica puramente scalare}\end{equation}

If the interferometer arm is parallel to the propagating GW, the longitudinal
response function, which has been obtained in \cite{key-2} with the
{}``bouncing photons analysis'' and in \cite{key-7} with a different
treatment, associated to such a massive mode is \begin{equation}
\begin{array}{c}
\Upsilon_{l}(\omega)=\frac{1}{m^{4}\omega^{2}L}(\frac{1}{2}(1+\exp[2i\omega L])m^{2}\omega^{2}L(m^{2}-2\omega^{2})+\\
\\-i\exp[2i\omega L]\omega^{2}\sqrt{-m^{2}+\omega^{2}}(4\omega^{2}+m^{2}(-1-iL\omega))+\\
\\+\omega^{2}\sqrt{-m^{2}+\omega^{2}}(-4i\omega^{2}+m^{2}(i+\omega L))+\\
\\+\exp[iL(\omega+\sqrt{-m^{2}+\omega^{2}})](m^{6}L+m^{4}\omega^{2}L+8i\omega^{4}\sqrt{-m^{2}+\omega^{2}}+\\
\\+m^{2}(-2L\omega^{4}-2i\omega^{2}\sqrt{-m^{2}+\omega^{2}}))+2\exp[i\omega L]\omega^{3}(-3m^{2}+4\omega^{2})\sin[\omega L]),\end{array}\label{eq: risposta totale lungo z massa}\end{equation}

where $m$ in eq. (\ref{eq: risposta totale lungo z massa}) is the
small mass of the particle associated to the GW and $v_{G}$ in eq.
(\ref{eq: metrica puramente scalare}) is the particle's velocity
(i.e. the group velocity as the massive GW has been analysed like
a wave-packet \cite{key-2,key-7}). The relation mass-velocity is
$m=\sqrt{(1-v_{G}^{2})}\omega$ \cite{key-2,key-7}. 

Thus, if advanced projects on the detection of GWs will improve their
sensitivity allowing to perform a GWs astronomy (we recall that signals
from GWs are quite weak) \cite{key-1}, one will only have to look
the interferometer response functions to understand if General Relativity
is the definitive theory of gravity. In fact, if only the two response
functions (\ref{eq: risposta totale Virgo +}) and (\ref{eq: risposta totale Virgo per})
will be present, one will conclude that General Relativity is definitive.
If the response function (\ref{eq: risposta totale Virgo scalar})
will be present too, one will conclude that massless Scalar - Tensor
Gravity is the correct theory of gravity. Finally, if a longitudinal
response function will be present, i.e. Eq. (\ref{eq: risposta totale lungo z massa})
for a wave propagating parallel to one interferometer arm, or its
generalization to angular dependences, one will learn that the correct
gravity theory will be massive Scalar - Tensor Gravity which is equivalent
to $f(R)$ theories. In any case, the analysed response functions
will represent the ultimate test for General Relativity. In fact,
General Relativity is the only gravity theory which admits only the
two response functions (\ref{eq: risposta totale Virgo +}) and (\ref{eq: risposta totale Virgo per})
\cite{key-4,key-7,key-17,key-18}. Such response functions correspond
to the two {}``canonical'' polarizations $h_{+}$ and $h_{\times}$
of standard General Relativity. Thus, if a third polarization will
be present, a third response function will be detected by GWs interferometers
and this fact will rule out General Relativity like the ultimate theory
of gravity. 

Resuming, in this proceeding paper we have shown that, by assuming
that advanced projects on the detection of GWs will improve their
sensitivity allowing to perform a GWs astronomy, accurate angular
and frequency dependent response functions of interferometers for
gravitational waves arising from various Theories of Gravity, i.e.
General Relativity and Extended Theories of Gravity, will be the ultimate
test for General Relativity.

\subsubsection*{Acknowledgements}

The Associazione Scientifica Galileo Galilei has to be thanked for
supporting this proceeding paper.


\begin{thebibliography}{31}
\bibitem{key-1}A. Giazotto - Journ. of Phys., Conf. Series 120, 032002
(2008) 

\bibitem[2]{key-2}C. Corda - J. Cosmol. Astropart. Phys. JCAP04009
(2007)

\bibitem[3]{key-3}\foreignlanguage{italian}{C. Corda - Int. Journ.}
Mod. Phys. A 23, 10, 1521-1535 (2008)

\bibitem[4]{key-4}Capozziello S and C. Corda - Int. J. Mod. Phys.
D \textbf{15,} 1119 -1150 (2006) 

\bibitem[5]{key-5}C. Corda - Astropart. Phys. 28, 2, 247-250 (2007)

\bibitem[6]{key-6}C. Corda - Astropart. Phys. 30, 4 209-215 (2008)

\bibitem[7]{key-7}S. Capozziello, C. Corda and M. F. De Laurentis
- Phys.Lett. B, 669, 5, 255-259, (2008) 

\bibitem[8]{key-8}T. Damour and G. Esposito-Farese - Class. Quant.
Grav. \textbf{9} 2093-2176 (1992);  

\bibitem[9]{key-9}\foreignlanguage{italian}{C. Corda - Int. Journ.}
Mod. Phys. A 22, 26, 4859-4881 (2007)

\bibitem[10]{key-10}S. Capozziello, C. Corda and M. F. De Laurentis
- Mod. Phys. Lett. A 22, 35, 2647-2655 (2007)

\bibitem[11]{key-11}S. Capozziello, C. Corda and M. F. De Laurentis
- Mod. Phys. Lett. A 22, 15, 1097-1104 (2007)

\bibitem[12]{key-12}C. Corda - Mod. Phys. Lett. A No. 22, 23, 1727-1735
(2007)

\bibitem[13]{key-13}C. Brans and R. H. Dicke - Phys. Rev. 124, 925
(1961)

\bibitem[14]{key-14}N. Bonasia and M. Gasperini - Phys. Rew. D \textbf{71}
104020 (2005)

\bibitem[15]{key-15}\foreignlanguage{italian}{C. W. Misner, K. S.
Thorne and J. A. Wheeler - {}``Gravitation'' - W.H.Feeman and Company
- 1973} 

\bibitem[16]{key-16}L. Landau and E. Lifsits - {}``Teoria dei campi''
- Editori riuniti edition III (1999) 

\bibitem[17]{key-17}S. Nojiri and S.D. Odintsov - Int. J. Geom. Meth.
Mod. Phys. 4, 115-146 (2007) 

\bibitem[18]{key-18}V. Faraoni and T. P. Sotiriou - arXiv:0805.1726,
to appear in Rev. Mod. Phys.; S. Capozziello and M. Francaviglia -
Gen. Rel. Grav. 40, 2-3, 357 (2008)

\bibitem[19]{key-19}A. Starobinsky - Phys. Lett. B, 91, 99-102 (1980)

\selectlanguage{italian}%
\bibitem[20]{key-20}\foreignlanguage{english}{P. J. E. Peebles and
B. Ratra - Rev. Mod. Phys. 75 8559 (2003) }

\selectlanguage{english}%
\bibitem[21]{key-21}C. M. - Will \textit{Theory and Experiments in
Gravitational Physics}, Cambridge Univ. Press Cambridge (1993)

\bibitem[22]{key-22}T. Inagaki, S. Nojiri, S. D. Odintsov - J. Cosmol.
Astropart. Phys., JCAP 0506 (2005) 010 

\bibitem[23]{key-23}http://www.aspera-eu.org/index.php?option=com\_content\&task=view\&id=125\&Itemid=99

\bibitem[24]{key-24}C. Corda - Astropart. Phys. \textbf{27,} No 6,
539-549 (2007)

\bibitem[25]{key-25}C. Corda - Int. J. Mod. Phys. D \textbf{16,}
9, 1497-1517  (2007)

\bibitem[26]{key-26}K. S. Thorne- \textit{300 Years of Gravitation}
- Ed. Hawking SW and Israel W Cambridge University Press p. 330 (1987)

\bibitem[27]{key-27}P. Saulson - \textit{Fundamental of Interferometric
Gravitational Waves Detectors} - World Scientific, Singapore (1994) 

\bibitem[28]{key-28} A. Pais - \textit{Subtle is the Lord} - Oxford
University Press (2005)

\bibitem[29]{key-29}C. Corda - \textit{Interferometric detection
of gravitational waves: the definitive test for General Relativity}
- Honorable Mention Winner at the 2009 Gravity Research Foundation
Awards for Essays on Gravitation, available on arXiv:0905.2502v1 {[}gr-qc{]}
15 May 2009

\bibitem[30]{key-30}http://www.aspera-eu.org

\bibitem[31]{key-31}Rakhmanov M - Phys. Rev. D \textbf{71} 084003
(2005) 
\end{thebibliography}
\end{document}